\documentclass{mem}
\usepackage{natbib}
\usepackage{graphicx}
\idline{00}{000}

\def\ltsima{$\; \buildrel < \over \sim \;$}
\def\lsim{\lower.5ex\hbox{\ltsima}}
\def\gtsima{$\; \buildrel > \over \sim \;$}
\def\gsim{\lower.5ex\hbox{\gtsima}}

\begin{document}
\def\teff{$T\rm_{eff }$}
\def\kms{$\mathrm {km s}^{-1}$}


\title{
INTEGRAL results on the electromagnetic counterparts of gravitational waves
}

   \subtitle{}



\author{
       S.~Mereghetti\inst{1}
            \and
   V.~Savchenko\inst{2}
	\and  
		C. Ferrigno\inst{2}
		\and 
		E.~Kuulkers\inst{3}
		\and
		P.~Ubertini\inst{4}
		\and
		A.~Bazzano\inst{4}
	\and
	E.~Bozzo\inst{2}
	\and 
	 S.~Brandt\inst{5}
	 \and
         J.~Chenevez\inst{5}
         \and
	  T.~J.-L.~Courvoisier\inst{2}
	  \and
		R.~Diehl\inst{6}
		\and 
		L.~Hanlon\inst{7}
		\and 
		A.~von~Kienlin\inst{6}
		\and
		P.~Laurent\inst{8,9}
		\and 
		F.~Lebrun\inst{9}
                \and
                A.~Lutovinov\inst{10,11}
                \and
                A.~Martin-Carrillo\inst{7}
     	        \and
                L.~Natalucci\inst{4}
		\and
		J.~P.~Roques\inst{12}
		\and
		T.~Siegert\inst{6}
                \and
                R.~Sunyaev\inst{10,13}
		}

\institute{
INAF, IASF-Milano, via E.Bassini 15, I-20133 Milano, Italy 
\and
ISDC, 
University of Geneva, chemin d'\'Ecogia, 16 CH-1290 Versoix, Switzerland 
\and
ESA/ESTEC, Keplerlaan 1, 2201 AZ Noordwijk, The Netherlands 
\and
INAF, IAPS,
Via Fosso del Cavaliere 100, 00133-Rome, Italy
\and
DTU,
Building 327, DK-2800 Kongens, Lyngby, Denmark 
\and
Max-Planck-Institut f\"{u}r Extraterrestrische Physik, Garching, Germany 
\and
University College Dublin, Belfield, Dublin 4, Ireland 
\and
APC, 
CNRS/IN2P3, CEA/Irfu, Observatoire de Paris
Sorbonne Paris Cit\'e, 10 rue Alice Domont et L\'eonie Duquet, 75205 Paris Cedex 13, France. 
\and
DSM/IRFU/SAp,
CEA Saclay, 91191 Gif-sur-Yvette Cedex, France 
\and
Space Research Institute,
Profsoyuznaya 84/32, 117997 Moscow, Russia  \label{SRI_russia}
\and
Moscow Institute of Physics and Technology, 
Dolgoprudny, 
141700, Russia
\and
Universit\'e Toulouse; 
IRAP; 9 Av. Roche, BP 44346, F-31028 Toulouse, France 
\and
MPI for Astrophysics, Karl-Schwarzschild-Str. 1, Garching
 D-85741, Germany 
}

\authorrunning{S.Mereghetti et al.}

\titlerunning{Electromagnetic counterparts of GW with INTEGRAL}

\abstract{
Thanks to its high orbit and a set of complementary detectors providing continuous coverage of the whole sky, the INTEGRAL satellite has unique capabilities for the identification and study of the electromagnetic radiation associated to gravitational waves signals and, more generally,  for multi-messenger astrophysics. Here we briefly review the results obtained during the first two observing runs of the advanced LIGO/Virgo interferometers.
}
\maketitle{}

\section{Introduction}

The INTEGRAL satellite, operating  since 2002, is the main mission of the European Space Agency devoted to observations in the hard  X-ray / soft $\gamma$-ray range with high spectral and angular resolution \citep{2003A&A...411L...1W}. 
A few unique properties   make it  a particularly powerful tool in the context of   multi-messenger astrophysics, that has recently  entered an exciting phase \citep{2017NatAs...1E..83V}, thanks to the high sensitivity reached by the LIGO/Virgo interferometers for gravitational waves (GW) and by the new generation of neutrino detetectors.  

The instruments on board INTEGRAL, besides providing  high sensitivity with good imaging and spectroscopic capabilities  over a wide field of view ($\sim$900 deg$^2$), are able to detect transient  $\gamma$-ray signals from every direction in the sky, as discussed in more detail in section \ref{sec_INT}.

The other crucial property of INTEGRAL in this context is its highly eccentric orbit, with a period of 2.7 days. This allows uninterrupted observations of virtually the whole sky for 85\% of the time (i.e. when the satellite is above the van Allen radiation belts). Note that, contrary to what happens for satellites in low earth orbits, the fraction of the sky occulted by the Earth is negligible (from 0.05\% at perigee to a maximum of $\sim$0.4\% when INTEGRAL is close to the radiation belts).  In addition, all the data are continuously transmitted to ground in real time and can be processed at the INTEGRAL Science Data Center \citep{2003A&A...411L..53C}, with a latency of only a few seconds from the time of their on-board acquisition \citep{2003A&A...411L.291M}.  

In the next sections we describe the performance of the INTEGRAL instruments and review the results obtained during the O1 (September 2015 -- January 2016) and O2 (December 2016 -- August 2017) observing runs of the LIGO/Virgo detectors.

\begin{figure}[t!]
\resizebox{\hsize}{!}{\includegraphics[clip=true]{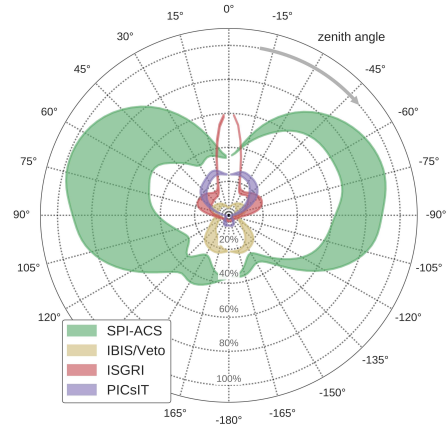}}
\caption{\footnotesize
Relative sensitivity of the  different detectors on board INTEGRAL as a function of the photon arrival direction \citep{2017A&A...603A..46S}. The shaded regions indicate the variation in sensitivity in the different azimuthal directions. A burst with duration of 1 s and Comptonized spectrum with $\alpha=-0.5$ and  $E_p$~=~600 keV has been assumed. 
}
\label{s1}
\end{figure}

\begin{figure}[t!]
\resizebox{\hsize}{!}{\includegraphics[clip=true]{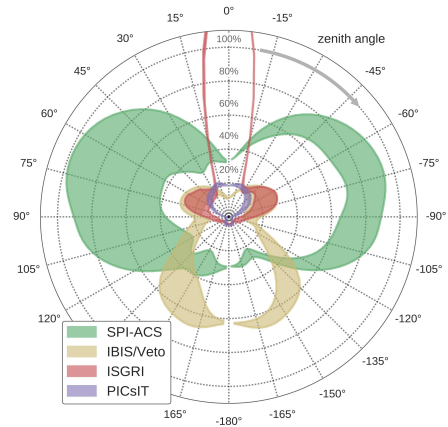}}
\caption{\footnotesize
Same as Fig.~\ref{s1}  but for a burst with duration of 8 s and a Band spectrum with   $\alpha=-1$, $\beta =-2.4$ and  $E_p$ = 300 keV.
Note that in this case the IBIS/VETO provides a better sensitivity than the SPI/ACS for events coming from the bottom direction.}
\label{s8}
\end{figure}

\section{INTEGRAL performances}
\label{sec_INT}

The INTEGRAL satellite carries two main instruments, SPI and IBIS,  operating at hard X-ray / soft $\gamma$-ray energies,  complemented by an X-ray and an optical telescope (JEM-X and OMC). All these instruments observe simultaneously and point in the same direction. IBIS uses two position-sensitive  detectors (ISGRI and PICsIt) coupled to a coded mask to provide images in the range from  20 keV  to 10 MeV over a  field of view of $\sim$30$^{\circ}\times30^{\circ}$ with an angular resolution of 12 arcmin \citep{2003A&A...411L.131U}. A similar field of view and energy range are covered by SPI. Its angular resolution is worse than that of IBIS, but thanks to its germanium detectors, SPI provides an excellent spectral resolution which makes it particularly useful to search for narrow lines from electron-positron annihilation or nuclear deexcitation \citep{2003A&A...411L..63V}.

Both the IBIS and SPI telescopes include active anticoincidence systems, based on BGO scintillators, that can be used very effectively as omnidirectional detectors capable to monitor  the entire  sky. Due to their different geometry and to the presence of surrounding absorbing material, the response of these detectors is a significant (and energy-dependent) function of the arrival direction of the photons (see Fig.~\ref{s1} and \ref{s8}). The highest sensitivity is given by the SPI anticoincidence shield (SPI/ACS), which is sensitive to photons of energy above $\sim$75 keV and provides light curves with fixed binning of 50 ms of the total count rate of the whole detector. 
The IBIS/Veto is sensitive in the 100 keV - 10 MeV range and provides light curves with a time resolution of 8 s.
Due to the lack of spectral and directional information of these detectors, whose main purpose is to shield the focal planes of the respective telescopes, it is necessary to assume a spectral shape and sky position  to convert their measured count rates to  photon fluxes in physical units.      
   
Finally, we note that, due to the high penetrating nature of $\gamma$-rays, both ISGRI and PICsIt are sensitive also to events coming from  sky regions outiside the imaging field of view. From a comparison of the relative number of  counts revealed in all the different elements that constitute the INTEGRAL payload it is possible to derive some rough information on the sky location of transient events.
 
A more complete description of the performances of the INTEGRAL instruments for the search of GW counterparts and other transient events can be found in \citet{2017A&A...603A..46S}.
 
\section{Results}

\subsection{GW 150914}
 
The first gravitational wave signal significantly detected during the O1 run of the Advanced LIGO interferometer, GW 150914, was located inside an uncertainty region (90\% confidence) with area of 630 deg$^2$ \citep{2016ApJ...826L..13A}. At the time of the GW trigger, INTEGRAL was pointing away from the   GW error region, but  its orientation was optimal to cover the whole uncertainty region with the SPI/ACS. 
Indeed in 95\% of the error region the achieved sensitivity was within 20\% of the best value, providing  constraining upper limits on the fluence  above 75 keV of possible counterparts  \citep{2016ApJ...820L..36S}. These limits depend on the assumed duration $\Delta$t  of the event, and to a lesser extent, on its sky position and spectral shape. For typical GRB spectra, the 3$\sigma$ upper limits range from $2\times10^{-8}$ erg cm$^{-2}$ ($\Delta$t=50 ms) to $\sim$10$^{-6}$ erg cm$^{-2}$  ($\Delta$t=10 s).

A possible hard X-ray transient lasting $\sim$1 s was detected with the Fermi/GBM instrument  about 0.4 s after the GW trigger time \citep{2016ApJ...826L...6C}. 
The significance of this event and its association to the GW source are subject of  discussion
\citep{2016A&A...593A..17G,2018arXiv180102305C}.
If confirmed, this would be a rather surprising result since GW 150914 was caused by the coalescence of two black holes \citep{2016PhRvL.116f1102A} and most models do not predict electromagnetic emission in this case.

A comparison of the Fermi/GBM results with the INTEGRAL upper limits is not straightforward, owing to the poorly constrained spectrum and uncertain arrival direction of this weak event. The GBM response extends to lower energies than that of the  SPI/ACS and, in principle, the results of the two instruments could be reconciled if the putative counterpart of  GW 150914 had a very soft spectral shape, different from that of the majority of GRBs. On the other hand, the GBM data favor a relatively hard spectrum (e.g. a Comptonized model with $\alpha_{comp} = -0.42$ and $E_{peak}>1$ MeV; \citet{2016ApJ...827L..34V}) that would result in a significant signal in the SPI/ACS.  Further work, also to investigate the relative intercalibration of the two instruments, is required to give a better assessment of the  properties of this electromagnetic signal and its possible association to GW 150914.

\subsection{GW 151226}
 
At the time of this event, produced by the coalescence of two black holes \citep{2016PhRvL.116x1103A}, INTEGRAL was not observing because it was close to the perigee, below the Earth radiation belts.

\subsection{GW 170104}  
   
GW 170104 was the first high-significance event revealed during the LIGO O2 observing run. Also this signal was caused by the merging of two black holes \citep{2017PhRvL.118v1101A}. 
It was localized within an uncertainty region (90\% confidence) of $\sim$1200 deg$^2$.  This region was entirely visible with good sensitivity by the SPI/ACS,  
but no significant signals were detected \citep{2017ApJ...846L..23S}. 
The derived upper limits are similar to those obtained for GW 150914.  For example, assuming a typical spectrum for a short GRB  (a cutoff power-law  with $\alpha=-0.5$ and  $E_p$ = 600 keV)  the SPI/ACS 3$\sigma$ upper limit for the 75-2000 keV fluence in a duration of 1 s is below 2$\times10^{-7}$ erg cm$^{-2}$ in 95\% of the LIGO localization region.

\citet{2017ApJ...847L..20V}  reported the possible detection of a weak and short (32 ms) burst, occurring 0.46 s before the GW trigger time, in the data of the MCAL detector on the AGILE satellite.  
For most of the localization region of GW 170104 the SPI/ACS provides an upper limit inconsistent with the fluence estimated with the AGILE/MCAL.

\begin{figure}[]
\resizebox{\hsize}{!}{\includegraphics[clip=true]{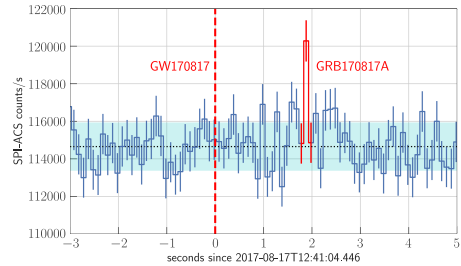}}
\caption{
\footnotesize
SPI/ACS light curve around the time of GW 170817, binned at 100 ms (from \citet{2017ApJ...846L..23S}).  The vertical dashed line indicates the time of the GW trigger.
}
\label{lc}
\end{figure}

\subsection{GW 170814}  
      
GW 170814 was the first event revealed by three gravitational waves interferometers. Thanks to the inclusion  of the Virgo data  it was possible to derive a small localization region of only 60 deg$^2$ \citep{2017PhRvL.119n1101A}. Also in this case, no significant signals were found with the SPI/ACS at or near the time of the GW trigger \citep{sav2017}.

An INTEGRAL  follow-up observation started about two days after the GW trigger and covered more than 90\% of the  localization region in the imaging field of view of IBIS and SPI with a maximum net exposure of $\sim$100 ks. No counterparts were found, with a 3$\sigma$ upper limit on the average flux of $\sim$3 mCrab (13 mCrab) in the 20--80 keV (80--300 keV) energy range \citep{sav2017e}.

\subsection{GW 170817}  
      
The first gravitational wave signal produced by the merger of two neutron stars was revealed by the the LIGO/Virgo interferometers on August 17, 2017 \citep{2017PhRvL.119p1101A}, while INTEGRAL was pointing toward the localization region of the previous event, GW 170814.  
The  independent discovery by the Fermi/GBM \citep{2017ApJ...848L..14G} and by the SPI/ACS \citep{2017ApJ...848L..15S} of an electromagnetic signal clearly associated to GW 170817 is a milestone of multi-messenger astrophysics. This event has   important physical and astrophysical  implications on many phenomena, such as, e.g.,  the speed of gravitational waves, the Lorentz invariance, the equivalence principle, the equation of state of neutron stars and the physics of GRBs \citep{2017ApJ...848L..12A}.

The SPI/ACS light curve around the time of GW 170817 is shown in Fig.~\ref{lc}. The excess corresponding to GRB 170817 is detected with a signal to noise ratio of 4.6,  1.9 s after the GW trigger time,  As  expected for such a faint $\gamma$-ray burst, no coincident signal was visible in all the other INTEGRAL detectors, thus supporting that the excess seen in the SPI/ACS was not due to particle background. We derived a 75-2000 keV  fluence of (1.4 $\pm$ 0.4 $\pm$ 0.6) $\times10^{-7}$ erg cm$^{-2}$, where the latter value gives the systematic error due to the uncertainty on the assumed spectral model.  

INTEGRAL carried out a follow-up observation, initially centered at the best Fermi/GBM location of GRB 170817, and later repointed toward the optical counterpart, as soon as it was announced.  This position was covered with a net exposure of more than 320 ks, starting about one day after the GW event,  but no X-ray or $\gamma$-ray counterparts were detected. 
The 3$\sigma$ upper limits on a long-lasting afterglow are of the order of $\sim$1--10   mCrab for energies below $\sim$100 keV and of few hundreds of mCrab in the MeV region.

Finally, thanks to  the  long INTEGRAL follow-up observation, we could also search for delayed bursting activity, as could be expected if (at least temporarily) a magnetar is formed, and for the presence $\gamma$-ray lines from r-process elements, such as I or Cs.
In both cases the results were negative, and no other mission could provide limits better than those obtained with the INTEGRAL instruments.

\section{Conclusions}

About fifteen years after its launch, INTEGRAL has  started a new exciting phase of its scientific life by playing a major role in the era of multi-messenger astrophysics.  In the case of black hole binary mergers, it has provided unique upper limits that constrain the ratio of emitted electromagnetic to gravitational energy to values E$_{\gamma}$/E$_{GW} \lsim10^{-7}-10^{-5}$. In the case of the first, and up to now single, GW event produced by the coalescence of two neutron stars, INTEGRAL has given a crucial independent confirmation of the short GRB discovered by Fermi/GBM, as well as important upper limits on subsequent high-energy  emission on different timescales.

The unique INTEGRAL performances discussed above are relevant also in the search for counterparts of astrophysical neutrinos, as demonstrated in  several recent cases for which constraining upper limits were provided \citep{manyGCNs,sant}.

The THESEUS satellite \citep{2017arXiv171004638A}, proposed for the ESA M5 call for new missions,  is planned to be operative in the years following 2030, when GW astronomy  will be a mature field, well beyond the current exploratory phase. The expected potentialities of THESEUS for multi-messenger astronomy are described in  \citet{2017arXiv171208153S}, but it is difficult to anticipate the wealth and variety of phenomena that THESEUS will address.  

The lesson that can be learned from the   INTEGRAL results described above, is that unanticipated uses of a payload can give important scientific contributions and exciting results. By definition, it is difficult to optimize the mission for an unforeseen science exploitation, but some general guidelines can be followed, as
including  the possibility of reconfiguration of the on-board software (with the associated problem of mantaining the required expertise for an extended time period). Also important are an accurate calibration of all the active elements (including  unconventional directions and energies), as well as a   complete characterization of both payload and spacecraft with an accurate mass model.


\bibliographystyle{aa}

\end{document}